\documentclass{emulateapj}
\usepackage{apjfonts,graphicx,ifpdf}
\input{epsf.sty}

\newcommand{\mum}{$\mu$m}
\newcommand{\HII}{H\,{\sc ii}}
\newcommand{\HI}{H\,{\sc i}}

\newcommand{\kms}{km~s$^{-1}$}
\def\hto{{\hbox{H$_{2}$O}}}
\def\cc{{\hbox{cm$^{-3}$}}}
\def\vh{{\hbox{V$_{\rm hel}$}}}

\shorttitle{Water Masers in Nearby Galaxies}
\shortauthors{Darling, Brogan, \& Johnson}

\begin{document}

\title{Ubiquitous Water Masers in Nearby Star-Forming Galaxies}

\author{Jeremy Darling\altaffilmark{1}, Crystal Brogan\altaffilmark{2}, \&
Kelsey Johnson\altaffilmark{3,4}}
\altaffiltext{1}{Center for Astrophysics and Space Astronomy,
Department of Astrophysical and Planetary Sciences,
University of Colorado, 389 UCB, Boulder, CO 80309-0389; 
jdarling@origins.colorado.edu}
\altaffiltext{2}{National Radio Astronomy Observatory, 520 Edgemont
  Rd, Charlottesville, VA 22903}
\altaffiltext{3}{Department of Astronomy, University of Virginia,
PO Box 3818, Charlottesville, VA  22903}
\altaffiltext{4}{KEJ is also an Adjunct Astronomer at the National Radio 
Astronomy Observatory}

\begin{abstract}

We report the detection of water maser emission from four nearby
galaxies hosting ultradense \HII\/ (UDHII) regions, He~2-10, the
Antennae galaxies (NGC 4038/4039), NGC 4214, and NGC 5253, with the
Green Bank Telescope.  Our detection rate is 100\%, and all of these
\hto\ ``kilomasers'' ($L_{\rm H_2O} <10$ L$_\odot$) are located
toward regions of known star formation as traced by UDHII regions and
bright 24 \mum\/ emission.  Some of the newly discovered \hto\/ masers
have luminosities 1--2 orders of magnitude less than previous
extragalactic studies and the same order of magnitude as those
typical of Galactic massive star-forming regions.  The unusual success
of this minisurvey suggests that \hto\ maser emission may be very
common in starburst galaxies, and the paucity of detections to date is
due to a lack of sufficient sensitivity. While the galaxy sample was
selected by the presence of UDHII regions, and the UDHII regions lie
within the telescope beam, in the absence of \hto\ spectral line maps
the connection between \hto\ masers and UDHII regions has not yet been
demonstrated.

\end{abstract}


\section{Introduction}

Extreme star-forming environments known as ``ultradense \HII\/
regions'' (UDHIIs) have been discovered in a number of starburst
galaxies \citep[e.g.,][]
{kobulnicky99,turner00,tarchi00,johnson04,reines08}.  
UDHIIs are identified through their
thermal mid-IR to radio emission (often being too deeply embedded in
their birth material to observe at optical wavelengths).  Similarly,
dense radio \HII\/ regions are known to exist on a much smaller scale
around individual massive stars in the Galaxy (i.e.,  ultracompact
\HII\/ regions [UCHIIs]; Wood \& Churchwell 1989).  However, UDHIIs
are vastly scaled up from individual UCHIIs, and they can contain the
equivalent of a thousand or more embedded O-type stars.

Given the tremendous number and density of young massive stars in
UDHIIs, it is logical to ask whether typical signposts of Galactic
massive star formation are associated with these objects.
\citet{churchwell90} found that $\sim 70$\% of Galactic UCHIIs are
associated with H$_2$O (22.235 GHz) masers.  Subsequent
interferometric follow-up showed that while some H$_2$O masers are
coincident with the UCHII regions themselves, many are instead
associated with younger members of the forming massive protocluster of
which the UCHII is a member \citep[e.g.,][]{tofani95, hofner96}. These
studies imply that the conditions required for \hto\/ masers ($n >
10^7$ \cc\/ and $T>400$ K) are fairly persistent during the early
stages of massive star cluster evolution.  If extragalactic UDHIIs are
composed of thousands of UCHIIs, then it is reasonable to expect
H$_2$O masers to be associated with a large fraction of the embedded
massive stars in these giant clusters.

The term ``kilomasers'' has been coined to describe extragalactic
22~GHz H$_2$O masers with luminosities comparable to the
brightest Galactic star formation-associated H$_2$O masers (e.g.,  W49N;
$L_{\rm H_2O}\sim 1$ L$_\odot$). The luminosities of H$_2$O kilomasers
($L_{\rm H_2O} <10$ L$_\odot$) are much lower than the more widely
studied ``megamasers,'' which are associated with the nuclear regions
of AGN (i.e.,  NGC~4258) and have been observed with luminosities up
to 10$^4$ L$_\odot$ \citep{barvainis05}. 
\hto\ kilomasers have been associated with both AGN and star formation
activity; this luminosity regime includes the tails (high and low) 
of both maser populations.
Although \hto\/ kilomasers can be used to help pinpoint sites
of active star formation, there have been few searches with the
sensitivity necessary to detect these faint masers at extragalactic
distances. To date six \hto\/ kilomasers have been unambiguously
associated with regions of known star formation in the LMC, M82,
IC~342, IC~10, M33 [IC~133], and NGC~2146 \citep[][and references
therein]{whiteoak86,tarchi02a,tarchi02b,henkel05,castangia08}.  The
six remaining known \hto\/ kilomasers either are associated with AGN
activity or have ambiguous provenance.  For example, the \hto\ masers
in NGC~253 appear to be associated with a nuclear outflow
\citep{hofner06} and not current star formation.

We have carried out a Green Bank Telescope\footnote{The National Radio
Astronomy Observatory is a facility of the National Science Foundation
operated under cooperative agreement by Associated Universities, Inc.}
(GBT) search for kilomasers toward four nearby starburst galaxies ($3<
D< 20$ Mpc), known to host UDHIIs, down to a sensitivity level
consistent with strong Galactic UCHII region \hto\/ masers. Positive
detections were found for all four galaxies as described in detail
below. The unusual success of this minisurvey is due to the fact that
our observations are more than an order of magnitude more sensitive
than the majority of previous single dish surveys for either
kilomasers or megamasers. This result suggests that \hto\/ maser
emission may be very common in starburst galaxies.

\section{Observations and Data Reduction}\label{sec:obs}

\begin{deluxetable*}{llcccrrcccl} 
\tabletypesize{\scriptsize}
\tablecaption{Journal of GBT Observations\label{tab:obs}}
\tablewidth{0pt}
\tablehead{
\colhead{Galaxy} &  
\colhead{Other Name} &  
\multicolumn{2}{c}{Pointing Center} &
\colhead{$S_{100\mu{\rm m}}$} &
\colhead{$D$} & 
\colhead{$t_{\rm int}$} & 
\colhead{$\Delta V_{\rm chan}$} &
\colhead{rms} &
\colhead{Observing} &
\colhead{UT Date} \\
\colhead{} & 
\colhead{} & 
\colhead{$\alpha$ (J2000)} & 
\colhead{$\delta$ (J2000)} & 
\colhead{(Jy)} &
\colhead{(Mpc)} &
\colhead{(h)} & 
\colhead{(km s$^{-1}$)} &
\colhead{(mJy)} & 
\colhead{Mode} & 
\colhead{} 
}
\startdata
He~2-10 &              & 08 36 15.3 & $-$26 24 30 
& 26.4     &
        10.5(0.7)\tablenotemark{a} & 2.2  & 3.3 & 1.2 & Nod & 2004 May 3, 4\\
Antennae&NGC 4038/4039 & 12 01 55.0 & $-$18 52 58 
& 76.0   &
        20.0(1.4)\tablenotemark{b} & 2.1 & 2.0 & 0.9 & OffOn & 2004 May 4, 6\\
NGC 4214 & UGC 7278    & 12 15 39.2 & +36 19 40   
& 25.5   &
        2.94(0.18)\tablenotemark{c} & 1.1 & 2.0 & 1.3 & OnOff & 2004 May 3\\
NGC 5253 & Haro 10     & 13 39 55.8 & $-$31 38 20 
& 29.8   & 
        3.33(0.17)\tablenotemark{d} & 2.9  & 0.66& 2.4 & Nod & 2004 May 3, 4
\enddata
\tablecomments{
Units of right ascension are hours, minutes, and seconds, and units 
of declination are degrees, arcminutes, and arcseconds.
The 100 \mum\/ flux densities come from the {\it IRAS} Point Source Catalog.
$t_{\rm int}$ indicates the total on-source integration time and
$\Delta V_{\rm chan}$ is the velocity resolution of spectra plotted in
Fig.\ \ref{fig:spectra}.}
\tablenotetext{a}{For He~2-10 we adopt an \HI\/ velocity-based Hubble flow distance that
accounts for local peculiar velocities \citep{mould00} and $H_\circ =
72(3)$~km~s$^{-1}$~Mpc$^{-1}$.}
\tablenotetext{b}{
The Antennae distance is controversial;
we adopt the conventional value of 19.2 Mpc  \citep{whitmore05} updated
to $H_\circ = 72(3)$~km~s$^{-1}$~Mpc$^{-1}$.
}
\tablenotetext{c}{The NGC 4214 distance is from the RG
branch method \citep{maiz02}.}
\tablenotetext{d}{The NGC 5253 distance is from Cepheids
\citep{gibson00}, and the uncertainty is statistical only (other
recent NGC~5253 Cepheid distances range from 3.0 to 4.1 Mpc).
}
\end{deluxetable*}

\begin{figure}
\epsscale{1.1}
\plotone{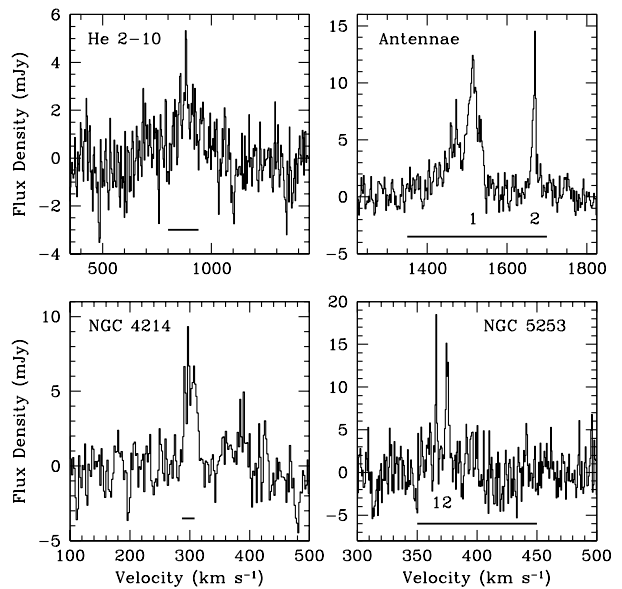}
\caption{Water maser spectra of 
He~2-10,
the Antennae galaxies (NGC 4038/4039),
NGC 4214, and
NGC 5253.  
Velocity resolutions (channel widths) are listed in Table \ref{tab:obs}.
Velocities are heliocentric.  
The bold horizontal bars indicate the velocity range of CO 
(see Table \ref{tab:lines} and \S \ref{sec:results}).  Numbers 
below the spectra label the lines enumerated in Table \ref{tab:lines}.
\label{fig:spectra}}
\end{figure}

We observed the $6_{16}-5_{23}$ 22.23508~GHz ortho-water maser line in
He~2-10, the Antennae (NGC 4038/4039), NGC 4214, and NGC 5253 with the
GBT on 2004 May 3--6 (Table \ref{tab:obs}). Two 25~MHz offset IFs
were observed simultaneously with 200~MHz bandpasses in two
polarizations and 12.2 or 24.4~kHz wide channels.  A winking
calibration diode and hourly atmospheric opacities were used for flux
density calibration.  Opacities ranged from 0.05 to 0.15 nepers but
were typically below 0.10.  The estimated uncertainty in the flux
density calibration is 10$\%$--15$\%$.  We used the 33\arcsec\ (FWHM) 
dual-beam {\it K}-band feed in a nodding mode for the compact galaxies (He~2-10
and NGC 5253) and a position-switched mode for the larger galaxies
(the Antennae and NGC 4214).  
Bandpasses were Hanning smoothed to 48.8~kHz channels for a rest-frame
velocity resolution of 0.66~km~s$^{-1}$ and Gaussian smoothed to the
resolution listed in Table \ref{tab:obs} to produce the final spectra
shown in Figure \ref{fig:spectra}.

For all observations, records were individually calibrated and
bandpasses flattened using the calibration diode and the corresponding
off-source records, respectively.  Scans and polarizations were
subsequently averaged, and a fifth-order polynomial baseline was fitted
and subtracted.  The bandpasses were rarely contaminated by radio
frequency interference (RFI) but were carefully inspected and
flagged.  Baselines were often not flat, but the atmospheric features
in the bandpass were much broader than any possible maser lines.  All
spectra were carefully inspected by polarization, IF, and time subsets
to confirm the validity of the detected lines.  All data reduction was
performed in GBTIDL, and all velocities are given in the
heliocentric frame.

\section{Results}\label{sec:results}

We have detected water maser emission in all four galaxies observed.  
Measured line properties are listed in Table \ref{tab:lines}, and
further information for each galaxy is presented below.
In all cases, given the complex velocity structure of the galaxies
in this sample (likely due to the combined effect of large and small
mergers and outflows), we might expect to find a correspondingly
complex velocity structure connected to the maser emission.  The
precise association of the maser emission with specific regions and
environments in these galaxies will only be possible with future 
high-resolution emission maps.

\begin{figure*}
\epsscale{1.15}
\plotone{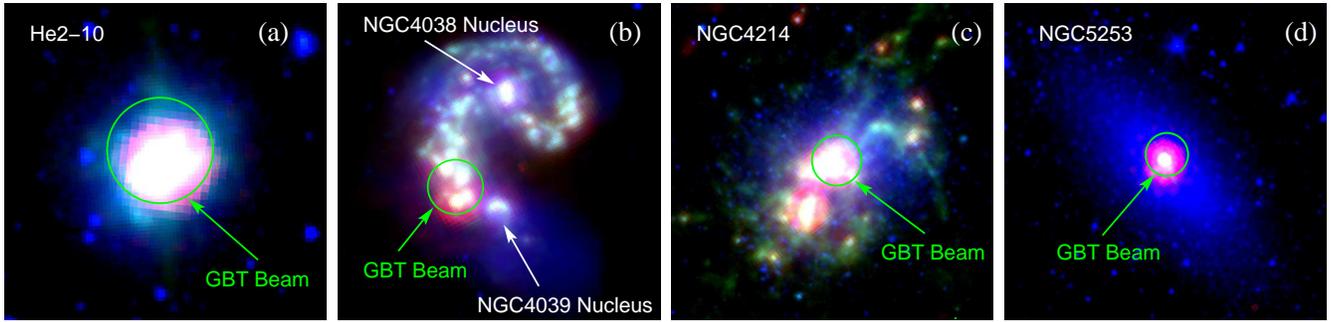}
\caption{Three-color archival {\em Spitzer} images of ({\it a}) He~2-10, ({\it b})
     the Antennae, ({\it c}) NGC~4214, and ({\it d}) NGC~5253 with RGB mapped to 24, 8,
     and 3.6 $\mu$m. In these images red colors pinpoint the most active
     regions of star formation. The locations of the $33''$ GBT beam are
     indicated with green circles; the positions of the two Antennae nuclei
     are also indicated.  The GBT beam corresponds to a physical scale in these
     galaxies of 1.7, 3.2, 0.47, and 0.53~kpc, respectively.
\label{fig:antennae}}
\end{figure*}

{\bf He~2-10:} \citet{kobulnicky99} identified five distinct UDHII regions
within the blue compact dwarf (BCD) galaxy He~2-10, which is located at
a distance of 10.5 Mpc.  He~2-10 has roughly solar metallicity
\citep[12+log(O/H)=8.93;][]{vacca92} implying that molecules may be
relatively abundant.  The He~2-10 UDHII regions are associated with
bright 24 \mum\/ emission, and the main body of the BCD fits
completely within the $33''$ GBT beam as shown in
Figure~\ref{fig:antennae}{\it a}, so at the current resolution it is not
possible to pinpoint where the \hto\/ emission originates.  The
velocity extent of CO(1--0) emission in He~2-10 observed with OVRO is
800--940 \kms\/ \citep{kobulnicky95}.  \citet{mohan01} find a
H92$\alpha$ line velocity peak of $\sim 950$~km~s$^{-1}$, with a broad
blue wing, and \citet{henry07} find narrow Brackett line features in
the range $\sim 860$--900~km~s$^{-1}$.  The maser velocity presented
here of 887~km~s$^{-1}$ is roughly consistent with all of these
studies.




{\bf The Antennae:} At an adopted distance of 20 Mpc,\footnote{
Recent work by \citet{saviane04,saviane08} suggests that the Antennae
distance may be as close as 13.3~Mpc.} the interacting galaxy pair
known as the Antennae (NGC~4038/NGC~4039) is the most distant source
in our survey.  As shown in Figure~\ref{fig:antennae}{\it b}, the single GBT
pointing ($33''$ beam) was centered on a 24 \mum\/-bright region of
copious star formation within the ``interaction region'' (IAR), offset
from both interacting galaxies' nuclei.  This region is also
coincident with several CO-identified ``supergiant molecular
complexes'' \citep[SGMCs;][]{wilson00} and UDHII regions identified by
VLA cm wavelength emission \citep{neff00}. The broad component of
maser emission that peaks at 1515~km~s$^{-1}$ is consistent with
velocities measured from ionized gas studies, such as
\citet{gilbert07}, as well as that of CO toward the IAR \citep[1350--1700 
\kms\/; see, e.g.,][]{wilson00,schulz07}. However, the
second maser component at 1670~km~s$^{-1}$ is on the high end of the
CO gas velocities observed within the IAR, and may instead be
associated with a ``bridge'' of CO at 1650--1700 \kms\/ that appears
to link the IAR to the NGC~4039 nucleus in the channel maps of
\citet{schulz07}. Emission from the NGC~4039 nuclear region itself is
unlikely because it lies near the first null of the GBT's beam; thus,
Antennae feature 2 is unlikely to be associated with nuclear activity.




{\bf NGC 4214:}
NGC~4214 is a BCD at a distance of 2.9 Mpc with a subsolar
metallicity \citep[12+log(O/H)=8.28;][]{kobulnicky96}. As shown in
Figure~\ref{fig:antennae}{\it c}, \hto\/ maser emission is detected toward a
region of bright 24 \mum\/ emission near the center of the galaxy. The
24 \mum\/ and \hto\/ emission are also coincident with several
unresolved thermal radio sources identified by \citet{beck00}. Using OVRO,
\citet{walter01} found that the CO(1--0) emission in this galaxy is
concentrated into three main regions, the centralmost of which is
coincident with the GBT pointing and the bright 24 \mum\/
emission. This CO emission has a peak \vh\/(CO)= 291.4 \kms\/ and a
velocity extent of 287--308 \kms, in good agreement with
\vh\/(\hto\/) = 301.6 \kms.



{\bf NGC 5253:} NGC~5253 is a dwarf starburst galaxy at a distance of
3.3 Mpc with subsolar metallicity
\citep[12+log(O/H)=8.23;][]{martin97}. Using OVRO, \citet{meier02}
found that the CO(2--1) emission is concentrated toward the optically
dark dust lane with a velocity range of 350--450 \kms. The GBT
beam was centered on the UDHII region discovered by \citet{turner00},
and a region of bright 24 \mum\/ emission (see
Fig.~\ref{fig:antennae}{\it d}). This region is also coincident with the
weak CO(2--1) molecular cloud "D" with \vh\/(CO)= 394 \kms\/ and
$V_{\rm FWHM}$=15 \kms\ \citep{meier02}.  However, the GBT beam also
encompasses parts of CO cloud "C" (\vh\/=418 \kms) to the southeast and
cloud "A" (\vh\/=363 \kms\/) to the east. Thus, the \hto\/ maser
emission at 366 and 375 \kms\/ is significantly blueward of the CO gas
toward the UDHII region and most closely agrees with the velocity of
cloud "A."  Using the radio recombinations lines H53$\alpha$ and
H92$\alpha$ \citet{mohan01} and \citet{rodriguez-rico07} measure peak
velocities of $\sim 400$~km~s$^{-1}$, both with broad wings.  These
velocities are somewhat higher than, but consistent with, the maser
velocities.
Given that the ionized gas in NGC~5253 is known to have a sinusoidal
velocity structure {\citep{lopez-sanchez07} with an amplitude of $\sim
20$~km~s$^{-1}$, it is not surprising that we see maser lines with
velocity separations within this range.

\begin{deluxetable*}{lccccccl}
\tabletypesize{\scriptsize} 
\tablecaption{Water Maser Line Properties\label{tab:lines}} 
\tablewidth{0pt} \tablehead{
\colhead{Galaxy} & \colhead{Line} & \colhead{$V_{H_2O}$} & 
\colhead{$V_{CO}$} &
\colhead{$S_{\rm Peak}$} & \colhead{$\Delta V_{\rm FWHM}$} & 
\colhead{$\int S{\rm d}v$} &
\colhead{$L_{H_2O}$} \\ 
\colhead{} & \colhead{} & \colhead{(km s$^{-1}$)} &
\colhead{(km s$^{-1}$)} &
\colhead{(mJy)} & \colhead{(km s$^{-1}$)} & \colhead{(mJy
km s$^{-1}$)} & \colhead{($L_\odot$)} } 
\startdata 
He~2-10  & 1 & 887(6) & 800--940 & 2.4(0.3) & 115(14) & 267(10) & 0.68(9) \\ 
Antennae& 1 & 1514.7(4)&  & 12.4(0.9) & 24(2)& 704(10)& \\ 
        & 2 & 1670.4(2)&  & 14.5(0.9) & 7(1) & 172(4) & \\ 
        & Sum & & 1350--1700 & & & 892(12) &  8.2(1.1)\\ 
NGC 4214& 1 & 301.6(8)& 287--308 & 6.7(0.6) & 21(2) & 141(6) & 0.028(4) \\ 
NGC 5253& 1 & 366.0(1)&  & 18.1(2.6)& 1.2(0.2) & 29(8) & \\ 
        & 2 & 375.1(2)&  & 13.2(1.6)& 3.3(0.5) & 47(9) & \\ 
        & Sum & & 350--450 & & & 83(13) &  0.021(4)
\enddata \tablecomments{
The line properties were determined from Gaussian fits, with the
exception of the Antennae, which were obtained directly from the data
because the line profiles are non-Gaussian.  The peak channel in the
He~2-10 spectrum is 4.5 $\sigma$ alone; the gaussian-fit peak is
$\sim$8 $\sigma$.  Isotropic line luminosities were computed from
$L_{H_2O} = (0.023\ L_\odot)\times D^2\times\int S{\rm d}v$, where $D$
is in Mpc and $\int S{\rm d}v$ is in Jy km s$^{-1}$ \citep{henkel05}.
$V_{CO}$ is the range of CO line velocities previously observed within
the GBT beam; references are provided in \S\ 3.}
\end{deluxetable*}


%

\section{Discussion}\label{sec:discussion}

Could the strong Galactic \hto\ maser source W49N be detected in our
sample galaxies?  Using a fiducial distance of 11.4~kpc and a 10~kJy
peak flux density \citep[see e.g.,][]{liljestrom89}, W49N is marginally
detectable in the Antennae (3~mJy) but easily detectable in the other
sample galaxies (12--150~mJy).  By luminosity ($\sim1$~$L_\odot$),
W49N should be detectable in NGC~4214 and NGC~5253 and possibly in
He~2-10, but not in the Antennae.  It is remarkable that we can now
detect sub-W49N H$_2$O masers in external galaxies, which means that
extragalactic \hto\ kilomaser observations may now be interpretable in
a Galactic star formation context.


Two of our new H$_2$O maser detections are extremely low luminosity; 
there is only one other system known with a total 
$L_{H_2O} < 0.1$~$L_\odot$, and this was a brief flaring 
episode in IC~342 \citep{tarchi02a,henkel05}.  The detection of 
deci-W49N emission in a larger sample of galaxies similar to NGC~4214
and NGC~5253 will provide a measurement of the \hto\ luminosity 
function at least an order of magnitude below current limits 
on $L_{H_2O}$ \citep{henkel05} and will genuinely explore the link
between \hto\ masers and star formation in galaxies as a function of 
environment, metallicity, and age.  

Our minisurvey sample has a 100\% detection rate, and expands by 2/3
the sample of known \hto\ kilomaser galaxies associated with star
formation.  The GBT beam excludes the nuclei in the Antennae, so the
association of \hto\ maser emission with star formation is likely in
this galaxy merger.  Moreover, the three dwarf galaxies in our sample
show no evidence of AGN activity.  Both He~2-10 and NGC~5253 have been
observed with the very long baseline interferometry (VLBI) 
High-Sensitivity Array (HSA) by
\citet{ulvestad07}, and neither has pointlike radio sources down to
a detection level of $1\sigma \sim 21~\mu$Jy, consistent with no AGN
activity.  The mid-infrared spectrum of NGC~4214 was studied by
\citet{satyapal08}, and they find no evidence of high-ionization lines
that would be suggestive of an AGN.  Indeed, given the association
between the presence of AGN and host galaxy bulge mass
\citep[e.g.,][]{kauffmann03}, it would be surprising if any of these
three dwarf galaxies harbored an AGN.  Without \hto\ spectral line
maps, however, the connection between \hto\ and UDHII regions has not
yet been demonstrated.

All four detected masers lie $\gtrsim1$ dex below the $\log (L_{H_2O}) =
\log (L_{FIR}) - 9.5$ relationship obtained for Galactic \hto\ masers by
\citet{felli92}.  That is, these masers are ``subluminous'' in the
\hto\ line given their host galaxies' far-IR luminosities.  They are,
however, in good agreement with the $\log L_{H_2O} = \log L_{FIR} -
10$ relationship for \hto\ kilomasers obtained by \citet{castangia08}.
The origin of this difference between kilomasers and Galactic masers
is unknown, but is likely due to spatially unresolved far-IR ({\it
IRAS}) observations of galaxies that include IR emission unrelated to
the sites of maser emission (Galactic sites of star formation were
much better resolved).  The difference between the \hto-FIR relation
of kilomasers and Galactic masers may thus not be physical at all
(unlike \hto\ megamasers, which represent a substantially different
physical process from Galactic and kilomasers).

\section{Conclusions}

We have identified H$_2$O masers in four galaxies hosting ultradense
\HII\ regions.  The high detection rate and the low luminosity of the
H$_2$O lines in our minisurvey indicate that previous water maser
surveys of nearby IR-bright galaxies lacked the sensitivity to detect
most H$_2$O kilomasers associated with bursts of star formation.  We
suggest that sensitive future surveys will produce a wealth of new
H$_2$O maser detections, enabling detailed studies of intense star
formation in other galaxies. Future interferometric imaging of the
maser lines detected with the GBT are needed to conclusively determine
whether the the maser emission is connected to massive star formation
in these galaxies. Indeed, although each of the GBT-detected masers
are likely composed of numerous individual maser spots, if any of them
remains relatively bright at very high spatial resolution (rather than
segregating into many weak spots), it may be possible to carry out
VLBI astrometry in the future as
has been recently done for several Local Group galaxies hosting \hto\
masers \citep{brunthaler07}.


\acknowledgements
This research has made use of the NASA/IPAC Extragalactic Database
(NED) and uses observations made with the {\it Spitzer Space Telescope},
both of which are operated by the Jet Propulsion Laboratory,
California Institute of Technology, under a contract with
NASA. K. E. J. acknowledges support from NSF through CAREER award
0548103 and the David and Lucile Packard Foundation through a Packard
Fellowship.  The authors would like to thank Jim Braatz for assistance
with calibration and data reduction and Francois Schweizer for helpful
discussions.


\clearpage

\end{document}